\documentclass{blois}

\def\be{\begin{equation}}
\def\ee{\end{equation}}
\def\bea{\begin{eqnarray}}
\def\eea{\end{eqnarray}}

\newcommand{\Photo}{}

\begin{document}
\vspace*{4cm}
\title{DOPED PARTON DISTRIBUTIONS}

\author{Valerio Bertone$^{1,2~}$\footnote{Speaker}, Stefano Carrazza$^3$ and Juan Rojo$^1$\\$\quad$\\}

\address{$^1$~Rudolf Peierls Centre for Theoretical Physics,\\
  1 Keble Road, University of Oxford, OX1 3NP, Oxford, UK\\
 $^2$ PH Department, TH Unit, CERN, CH-1211 Geneva 23, Switzerland\\  
 $^3$~Dipartimento di Fisica, Universit\`a di Milano and
INFN, Sezione di Milano,\\ Via Celoria 16, I-20133 Milano, Italy
}

\maketitle\abstracts{
  Calculations of high-energy processes involving the production of $b$ quarks
  are typically performed in two different ways, the massive four-flavour scheme (4FS) and
  the massless five-flavour scheme (5FS).
  For processes where the combination of the 4FS and 5FS results into
  a matched calculation is technically difficult, it
  is possible to define a hybrid scheme known as the {\it doped} scheme,
  where above the $b$-quark threshold the strong coupling runs with $n_f=5$, as in
  the massless calculation, while the DGLAP splitting functions are those of the
  $n_f=4$ scheme.
  In this contribution we present NNPDF3.0 PDF sets in this { doped} scheme,
  compare them with the corresponding 4FS and 5FS sets, and discuss their
  relevance for LHC phenomenology.}

\paragraph{Motivation}
The accurate modeling of heavy quark mass effects is a crucial
ingredient of LHC phenomenology.
In processes involving $b$-quarks, two calculational schemes are usually employed:
the massless $n_f=5$ scheme (5FS), where the $b$-quark mass is neglected and potentially large
logarithms of the type $\ln Q^2/m_b^2$ are resummed into a $b$-quark PDF $b(x,Q^2)$, and the
massive $n_f=4$ scheme (4FS),
where the non-zero value of $m_b$ is kept in the calculation.
Each scheme has its own advantages and weakness.

A general framework
for the combination of 4FS and 5FS results into a single  matched calculation, FONLL, is available,
and has been applied to several processes from heavy quark hadroproduction~\cite{Cacciari:1998it} to
DIS structure functions~\cite{Forte:2010ta} and Higgs production via bottom quark fusion~\cite{Forte:2015hba}.
The implementation of FONLL is however technically challenging for
processes for which analytic expressions are not available (such
as in semi-automated NLO codes) and therefore a hybrid scheme that captures the essence
of this matching may be useful, even though it will
be necessarily less accurate than the full matched calculation.

As discussed in Ref.~\cite{Maltoni:2012pa}, for many LHC processes the 4FS
is perfectly adequate, since potentially large collinear logarithms in $m_b$ are suppressed both
by the smallness of the $b$-PDF at large-$x$ and by universal phase-space factors.
However, a known drawback of this scheme is that with the 4FS evolution of the $\beta$ function
for the running of the strong coupling, above the $b$-quark
threshold $Q=m_b$ the value of $\alpha_s(Q)$ is not accurately reproduced,
and therefore the strong coupling running
cannot be simultaneously correct at relatively low scales, relevant for many of the
datasets that enter
global PDF fits, and at the higher scales relevant for the majority of LHC processes.
This difference induces
a suppression in the predicted cross-sections, which can be especially important
for multi-leg processes at the Born level where the calculation starts with a high power
of $\alpha_s$.

To overcome these difficulties, a hybrid scheme for the treatment of heavy quark mass effects
has been proposed~\cite{Napoletano}, named the {\it doped} scheme, where $\alpha_s(Q)$ runs
with $n_f=5$ flavors above $m_b$, thus having the standard value at $M_Z$
(consistent with the global PDG average), while DGLAP
evolution is still performed in the $n_f=4$ scheme, and in particular no $b$-quark
PDF is introduced, as in the massive scheme.
This scheme allows the combination of the useful properties of the 4FS, in particular
accounting for bottom quark mass effects, while also using a value of $\alpha_s(Q)$ consistent
with the global average for all values of $Q$.

Therefore, the basic idea of the doped scheme is to use, above the $b$-quark threshold,
a four-flavor factorization scheme (therefore a 4FS for the DGLAP evolution) with a
five-flavor renormalization scheme (thus, a 5FS for the running of $\alpha_s(Q)$).
However, as explained in Ref.~\cite{Martin:2006qz},
this choice is inconsistent, because
then there would be a mismatch between terms in the splitting functions
(specifically $P_{gg}$) which depend on the beta function, and those
governing the evolution of the strong coupling.
Recently, it was suggested in Refs.~\cite{Napoletano,Cascioli:2013era}
that this inconsistency can be cured by subtracting
order by order from the partonic matrix
elements the difference between the 4FS and 5FS running of $\alpha_s(Q)$, up to the order
at which the hard cross-section is computed.
Once this subtraction is implemented,
one consistently restores the 4FS up to the finite order at which the
matrix element has been computed.

\paragraph{NNPDF3.0 sets in the doped scheme}

We now present results for a version of
the NNPDF3.0 NLO and NNLO global PDF sets~\cite{Ball:2014uwa} that are
suitable to be used the doped
scheme calculations,
and compare them with their 4FS and 5FS counterparts.
Following the standard method adopted in the NNPDF framework~\cite{Ball:2011mu}, the doped sets are constructed starting from the
NNPDF3.0 $n_f=5$ global fit at a scale equal to the $b$-quark mass, $Q^2=m_b^2$, and
then evolving upwards in $Q^2$ using the DGLAP equations but now with $n_f=5$ in the $\beta$-function
for the running of $\alpha_s(Q^2)$ and with $n_f=4$ in the splitting functions.
The DGLAP upwards evolution is performed with the {\tt APFEL} program~\cite{Bertone:2013vaa},
using the same truncated
solution of the evolution equations as that used in the NNPDF3.0 fits.

\begin{figure}[t]
  \centering
   \includegraphics[scale=.34]{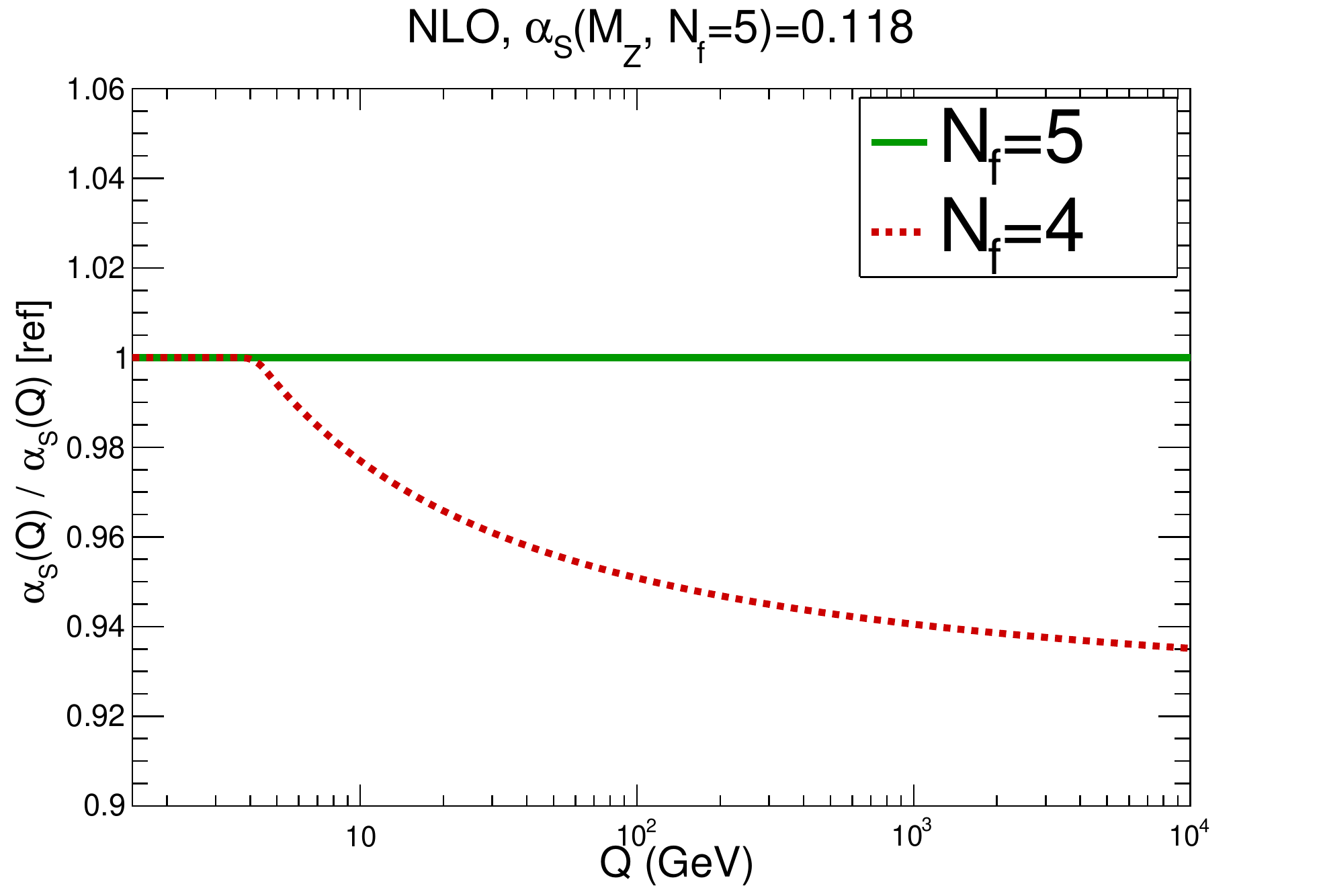}
  \includegraphics[scale=.34]{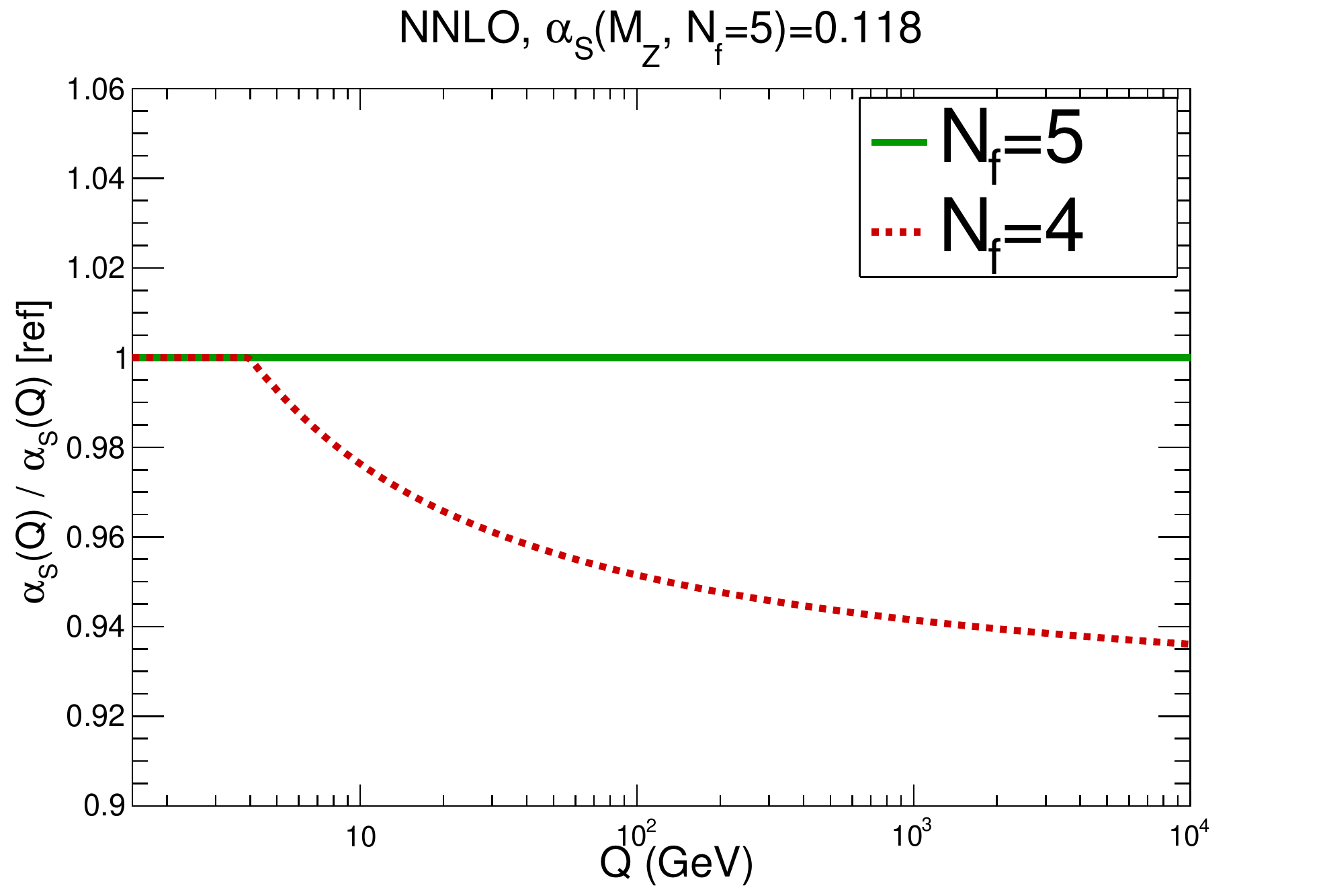}\\
  \caption{\small Running of the strong coupling $\alpha_s(Q)$
    in the 5FS and 4FS, normalized to the 5FS result.
  }  
\label{fig:as}
\end{figure}

First of all, we show in Fig.~\ref{fig:as} a comparison of the running of
the strong coupling $\alpha_s(Q)$ in the 5F and 4F schemes.
The different running in the 4F as compared to the 5F
scheme induces variations on the value of $\alpha_s(Q)$ between
3\% at $Q\sim 10$ GeV, and 6\% at $Q\sim 10$ TeV,
which in turns would lead to a shift $\delta \sigma/\sigma \sim n \delta \alpha_s/\alpha_s$
for processes whose 4FS calculation starts as $\sigma \sim \alpha_s^n$ at the Born level.

%%%%%%%%%%%%
%%%%%%%%%%%%%%%%%%%%%%%%%%%%%%%%%%

\begin{figure}[t]
  \centering
   \includegraphics[scale=.35]{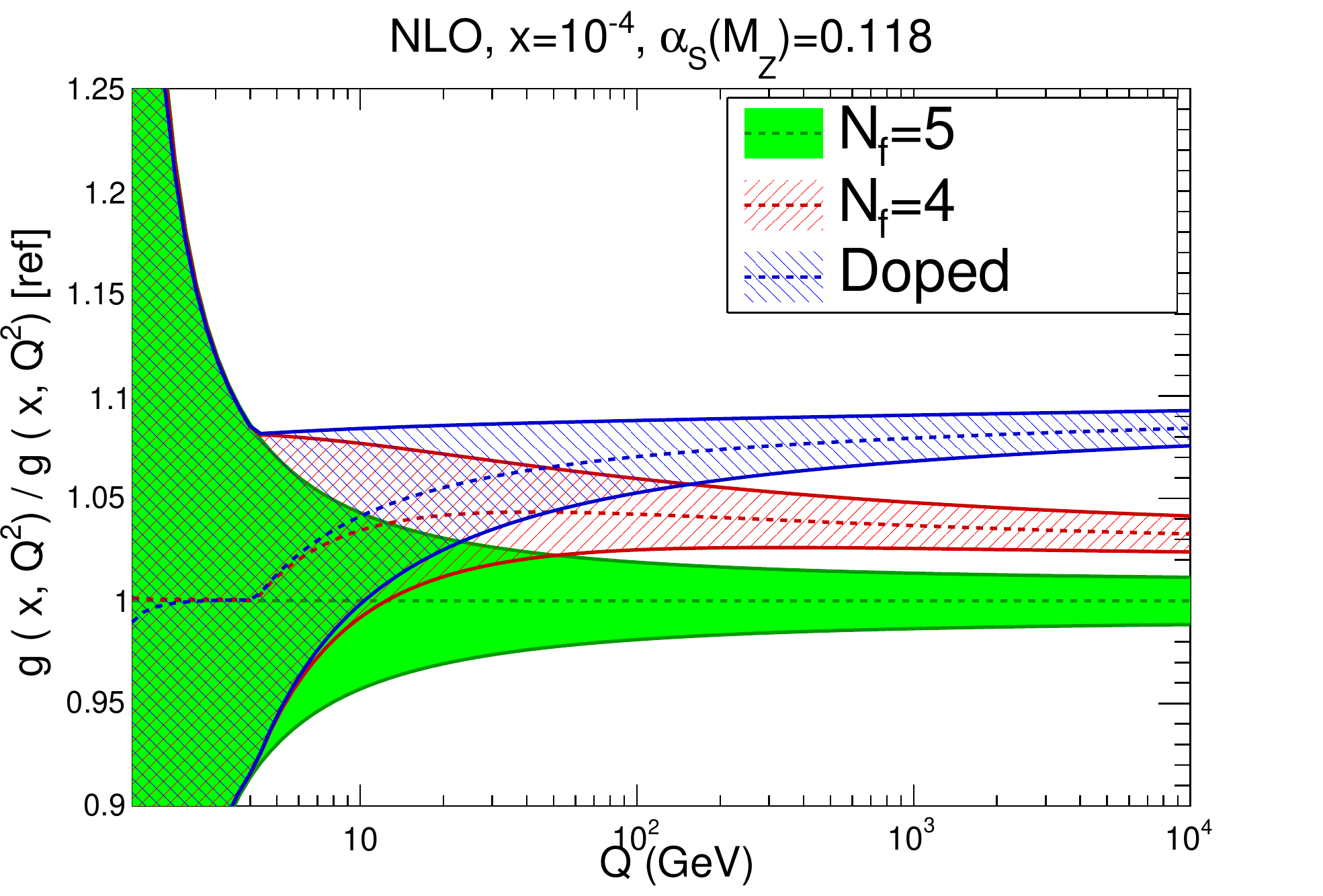}
  \includegraphics[scale=.35]{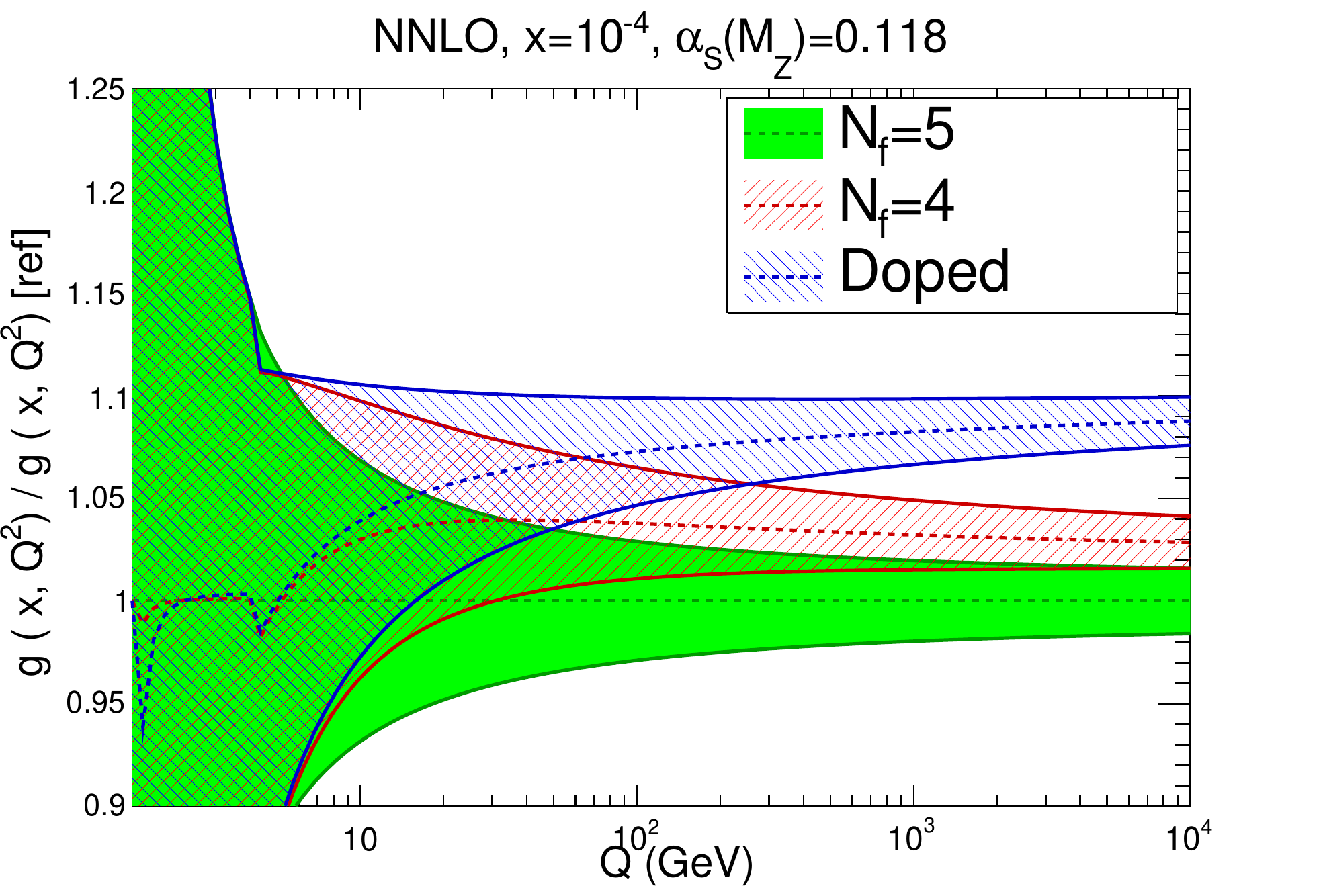}\\
   \includegraphics[scale=.35]{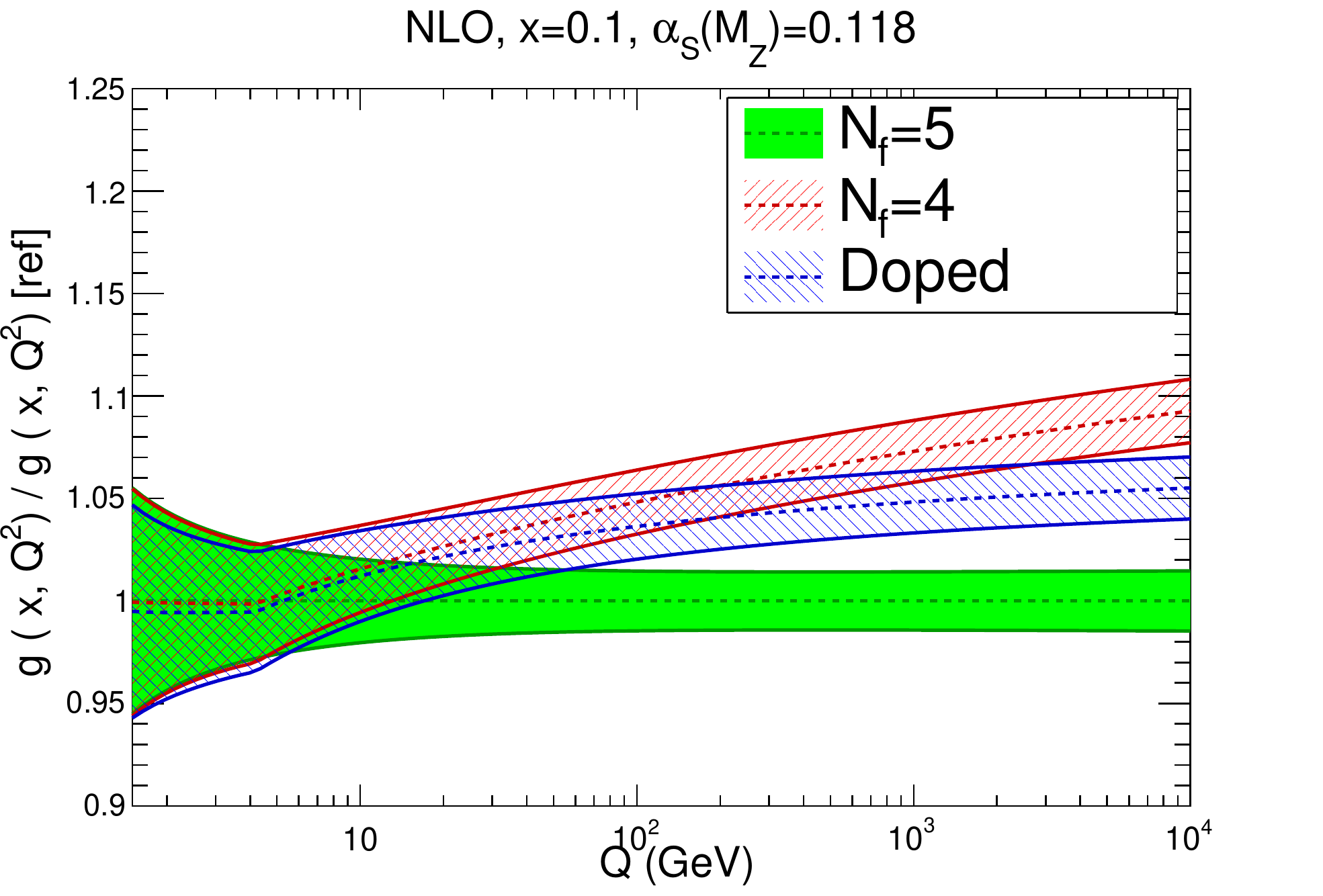}
   \includegraphics[scale=.35]{plots/xg-doped-nlo-largex.pdf}\\
   \caption{\small Comparison of the NNPDF3.0 global fits in the
     5F, 4F and doped heavy quark schemes, as a function of
     the factorization scale $Q$ for two different values of $x$, $x=10^{-4}$ (upper plots)
     and $x=0.1$ (lower plots).
     We show results normalized to the central value of the 5FS PDF set, for both
     NLO (left plots) and NNLO (right plots).
  }  
\label{fig:xg}
\end{figure}

\begin{figure}[t]
  \centering
   \includegraphics[scale=.37]{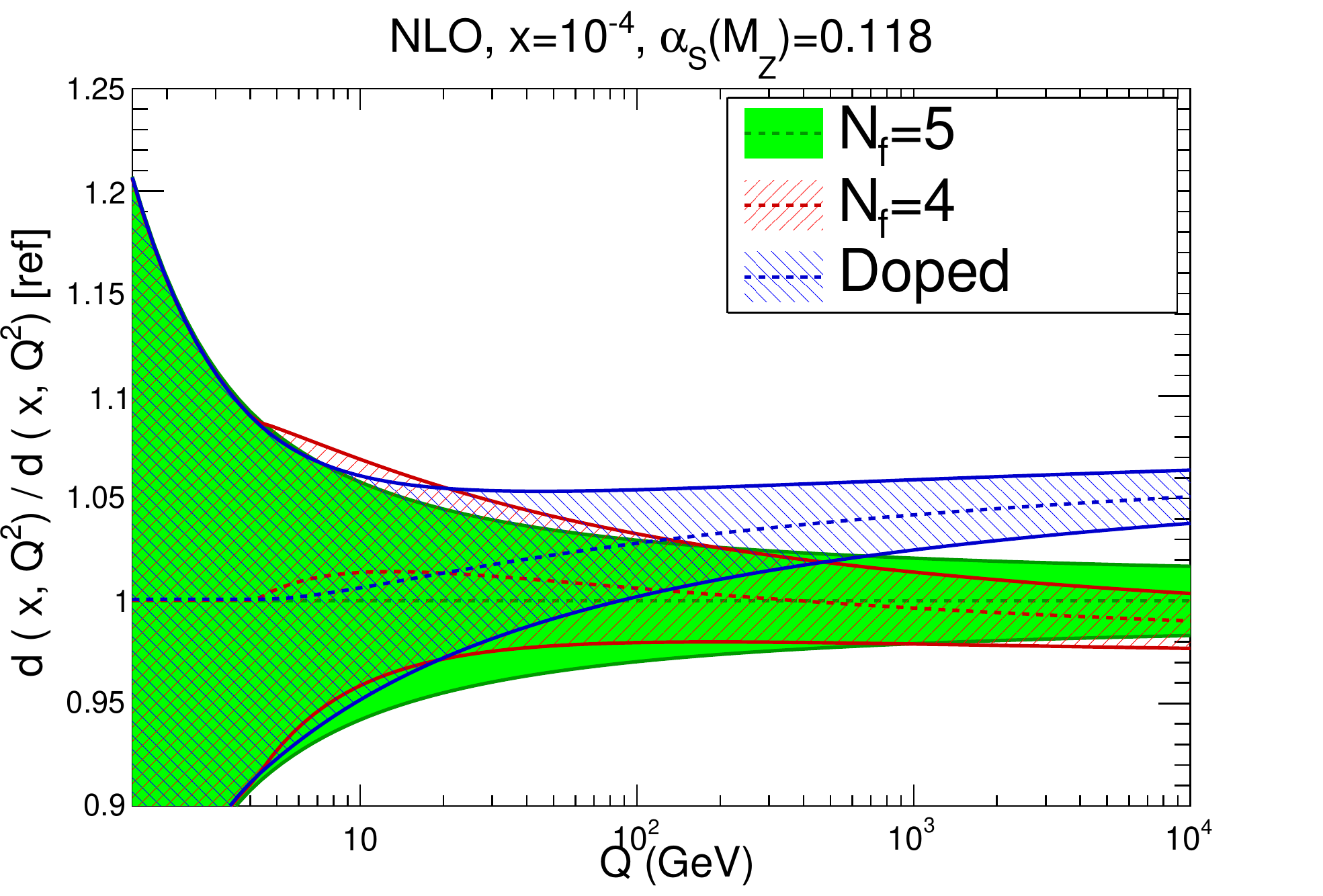}
   \includegraphics[scale=.37]{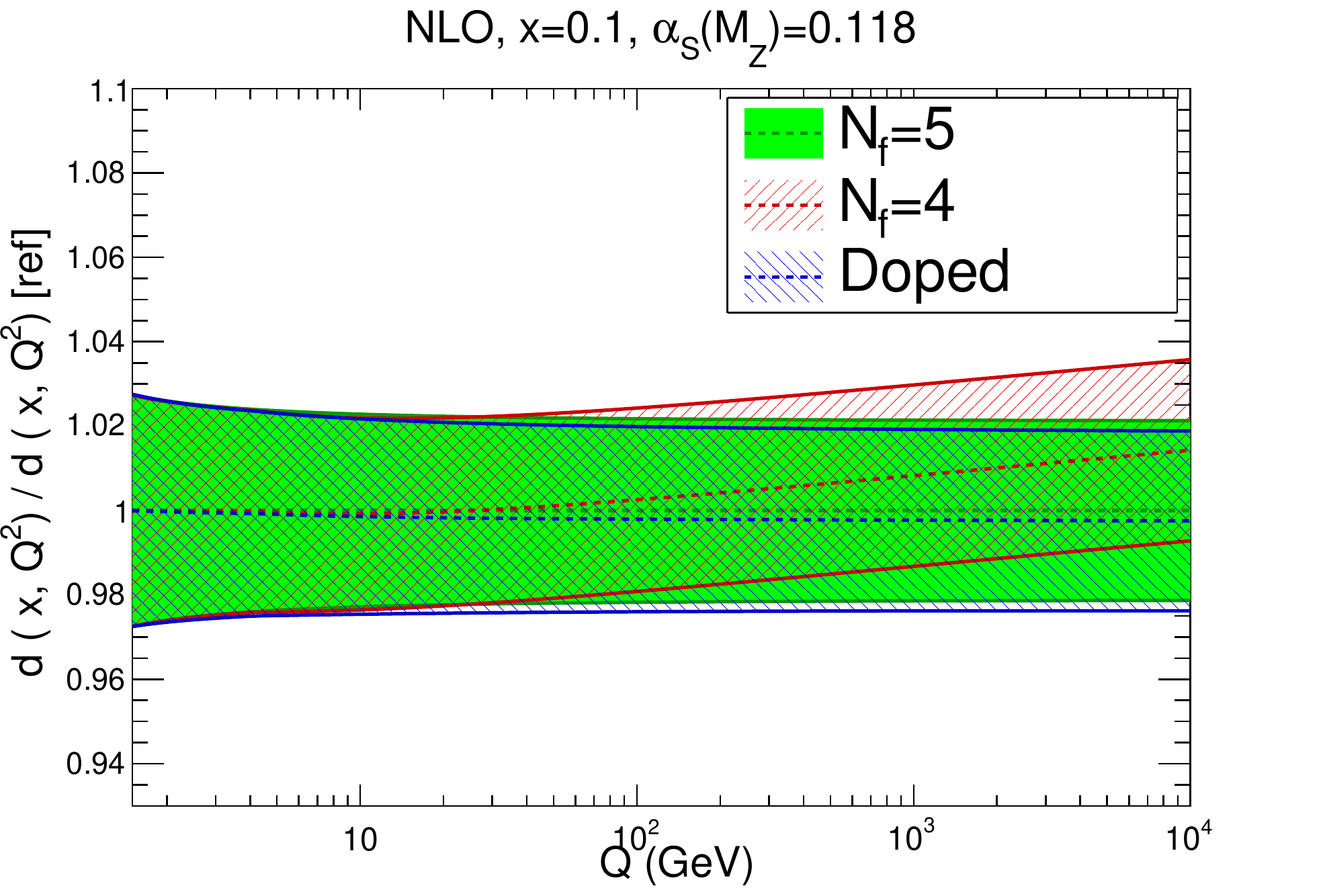}
   \caption{\small Same as Fig.~\ref{fig:xg} for the NLO down quark PDF.
  }  
\label{fig:xd}
\end{figure}

In Fig.~\ref{fig:xg} we compare the gluon PDF, as function of $Q$, for two values of $x$,
$x=10^{-4}$ and $x=0.1$, at NLO and at NNLO, between the 5F (reference),
4F and doped schemes.
As can be seen from the comparison, at large-$x$ the doped scheme leads to
intermediate results between the 4F and 5F schemes.
On the other hand, at small-$x$, the doped scheme leads to a stronger rise of
the gluon at small-$x$ as compared to the 4F calculation, due to the larger value
of $\alpha_s(Q)$ at high scales, which drives the small-$x$ DGLAP evolution.
The results are qualitatively the same at both perturbative orders, NLO and NNLO.
The same comparison, now for the down quark PDF $d(x,Q^2)$ is shown in
Fig.~\ref{fig:xd}.
At large values of $x$, where effects of DGLAP evolution are moderate, the doped
scheme coincides essentially with the 5F scheme, and the differences with the 4FS
are a few percent at most.
At small-$x$, the doped scheme also leads to a steeper rise of the light
quark sea.

%%%%%%%%%%%%
%%%%%%%%%%%%%%%%%%%%%%%%%%%%%%%%%%

%
%%%%%%%%%%%%
%%%%%%%%%%%%%%%%%%%%%%%%%%%%%%%%%%

\paragraph{Implications for LHC phenomenology}

\begin{figure}[t]
  \centering
   \includegraphics[scale=.37]{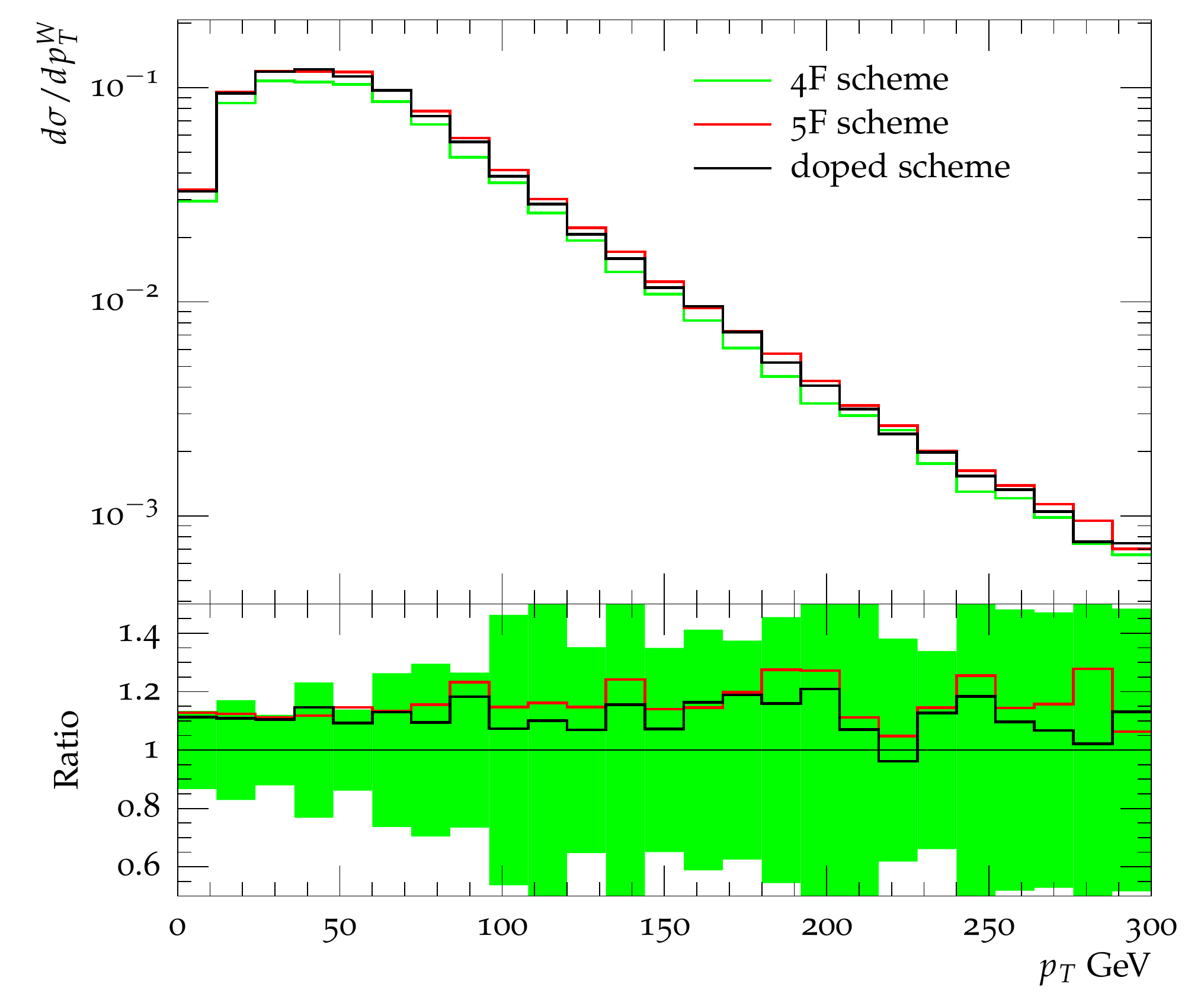}
  \includegraphics[scale=.37]{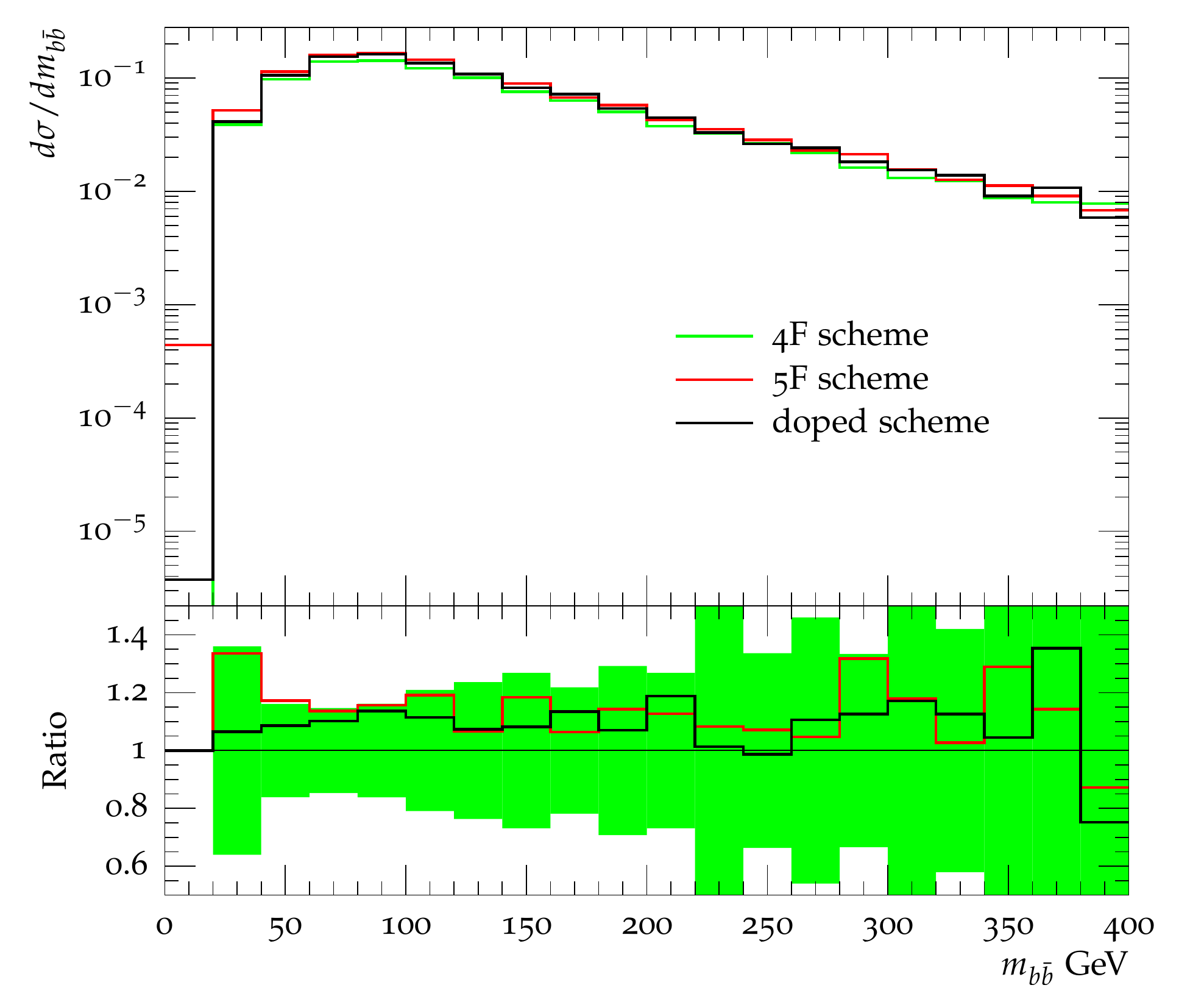}
  \caption{\small \label{fig:LHCVbb}
    Differential distributions for vector boson production 
    with a pair of $b$-quarks at the LHC 7 TeV: the $p_T^W$ of the $W$ boson
    in $Wb\bar{b}$ production (left) and the invariant mass of the
    $b\bar{b}$ pair in $Zb\bar{b}$ production (right).
    The results of the 4F and 5F schemes are compared with the doped scheme.
    }  
\end{figure}

A first study of the use of the doped scheme at the LHC for
the production of vector bosons associated with a $b\bar{b}$ pair
was presented in Ref.~\cite{Napoletano}.
As an illustration of these results, in Fig.~\ref{fig:LHCVbb}
we show differential distributions for
$Vb\bar{b}$ production at the
LHC for $\sqrt{s}=7$ TeV, in particular the the $p_T^W$ of the $W$ boson
    in $Wb\bar{b}$ production and the invariant mass of the
    $b\bar{b}$ pair, $m_{b\bar{b}}$ in $Zb\bar{b}$ production.

    These results were obtained in the framework
    of the {\tt Sherpa}+{\tt OpenLoops}~\cite{Gleisberg:2008ta,Cascioli:2011va} Monte Carlo generator,
    using the 5FS, 4FS and doped versions
    of NNPDF2.3 NLO~\cite{Ball:2012cx}.
    In each case, a consistent treatment of the partonic matrix elements was adopted.
    From Fig.~\ref{fig:LHCVbb} we see that the differential distributions
    in the doped scheme are typically intermediate between the results
    from the 4F and 5F schemes, as one could expect from the
    comparison at the level of PDFs shown in Figs.~\ref{fig:xg} and~\ref{fig:xd}.
    These comparisons imply that the (resummed) change in $\alpha_s(Q)$ and
    in the PDFs is more significant than the finite-order correction in the
    matrix element implemented in the doped scheme.
    The results of the calculations in the three schemes are consistent
    with the scale uncertainties of the 5FS calculation.

    Doped PDFs were also used in the $t\bar{t}b\bar{b}$ calculation of
    Ref.~\cite{Cascioli:2013era}, where it was found that
    a resummation of the logarithms of $\mu_R/m_b$ arising from the
    running of $\alpha_s$ in the doped scheme increased to
    total NLO cross-section by almost 10\% as compared to the 4FS calculation.

%%%%%%%%%%%%%%%%%%%%%%%%%%%%%%%%
%%%%%%%%%%%%%%%%%%%%%%%%%%%%%%%%%%

\begin{figure}[t]
  \centering
  \includegraphics[scale=.65]{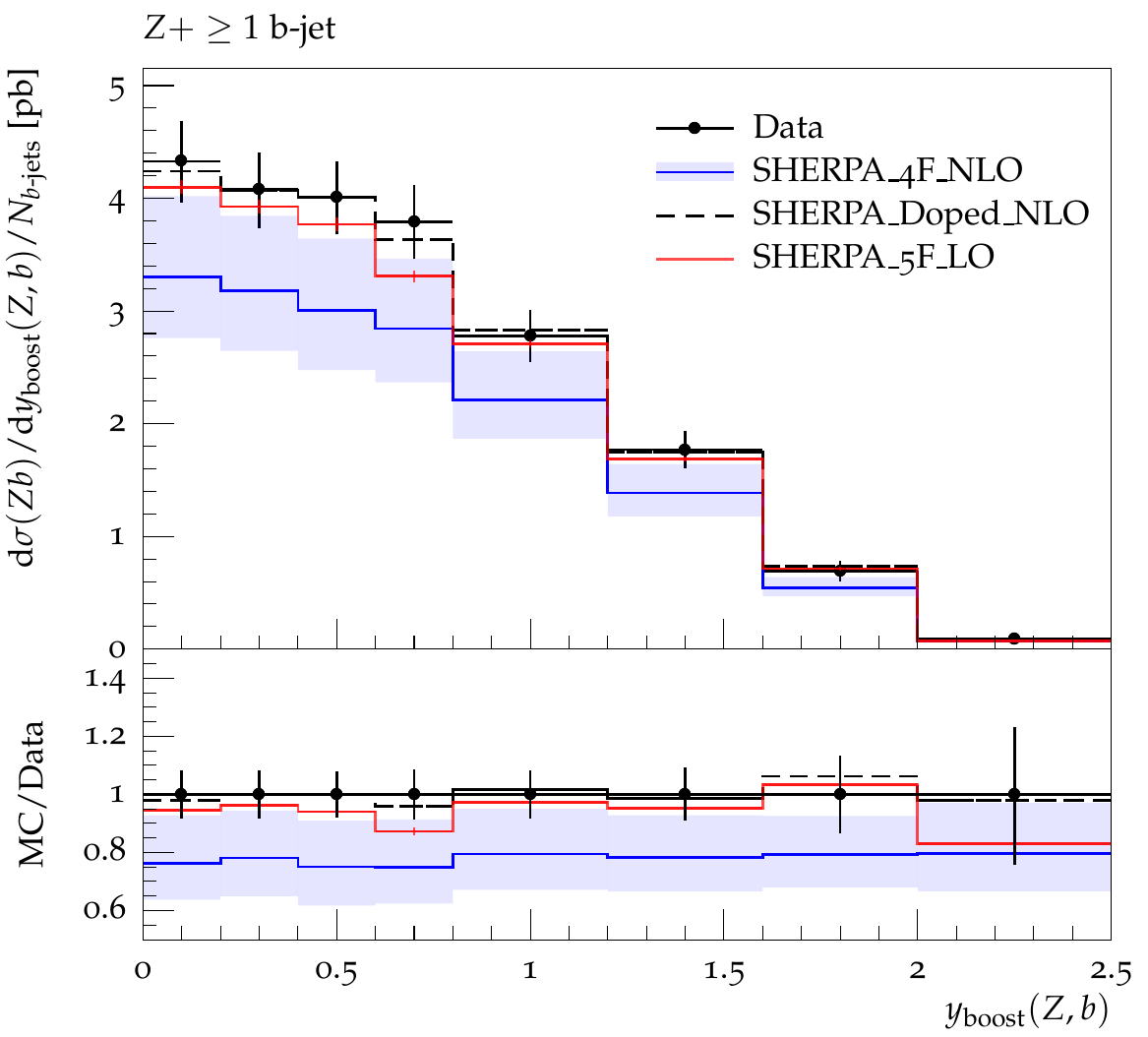}
  \includegraphics[scale=.65]{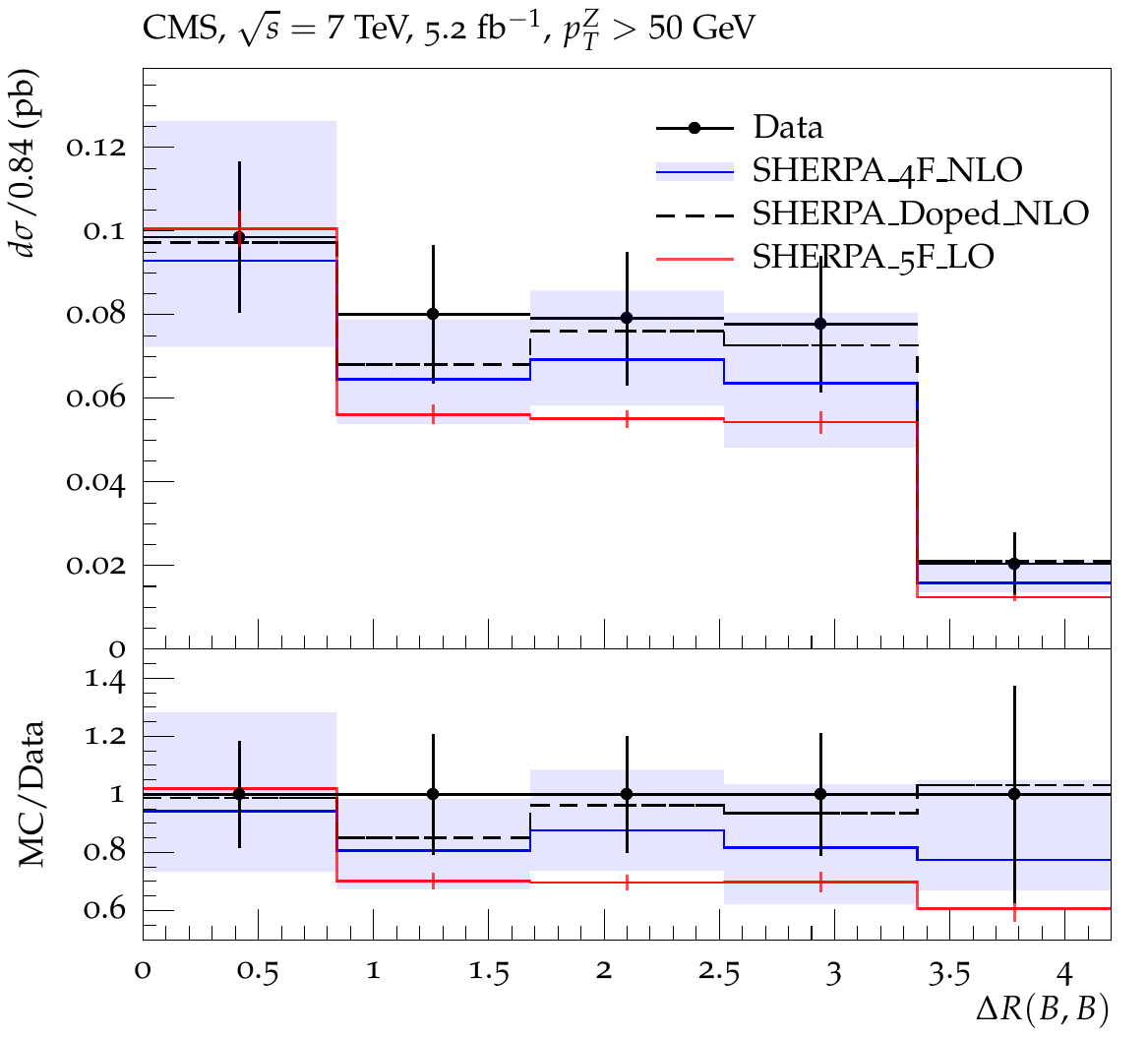}
  \caption{\small \label{fig:LHCVbbData}
    Preliminary results
    for the comparison between the 4FS NLO, doped NLO and 5FS LO
    calculations with {\tt Sherpa} using the corresponding NNPDF3.0 PDFs,
    and differential
    distributions from the recent ATLAS and CMS 7 TeV measurements of $Z$ production
    in association with $b$-jets.
    }  
\end{figure}

The NNPDF3.0 doped PDFs presented in this contribution
have been recently used~\cite{davide} in the {\tt Sherpa}
framework to compare the predictions of the 4F, 5F and doped schemes with
recent
ATLAS~\cite{Aad:2014dvb} and CMS~\cite{Chatrchyan:2013zja} measurements
at 7 TeV for the differential distributions of
$Z$ bosons produced in association with $b$-jets.\footnote{We thank Davide Napoletano for providing
  these comparison plots.}
These comparisons are illustrated in Fig.~\ref{fig:LHCVbbData}, showing that in
general the doped scheme leads to a better agreement with the LHC data as compared to
the 4F and 5F schemes; for instance for the ATLAS $y_{\rm boost}(Z,b)$ distribution, the
doped scheme leads to the correct normalization and shape while the 4FS reproduces
the shape but undershoots the data.

%%%%%%%%%%%%%%%%%%%%%%%%%%%%%%%%
%%%%%%%%%%%%%%%%%%%%%%%%%%%%%%%%%%

    Summarizing, the doped scheme is a hybrid scheme for the treatment of heavy quark masses
    in perturbative QCD calculations, which has the potential to improve some
    of the drawbacks of 4FS calculations, in particular for processes with many colored
    partons in the final state.
    The NLO and NNLO doped versions of NNPDF3.0 will be made available
    in {\tt LHAPDF6}.

\end{document}